\documentclass[aps,prl,
twocolumn,
,showpacs,superscriptaddress]{revtex4}  
\usepackage{graphicx,dcolumn,bm,amsmath,amssymb} 
\renewcommand{\vec}[1]{\boldsymbol{#1}}
\begin{document}

\title{Radiation friction vs ponderomotive effect}
\author{A.M.~Fedotov}\affiliation{National Research Nuclear University ``MEPhI'' (Moscow Engineering Physics Institute), 115409 Moscow, Russia}
\author{N.V.~Elkina}\affiliation{Ludwig-Maximilians Universit\"{a}t, M\"unchen, 80539, Germany}
\author{E.G.~Gelfer}\affiliation{National Research Nuclear University ``MEPhI'' (Moscow Engineering Physics Institute), 115409 Moscow, Russia}
\author{N.B.~Narozhny}\affiliation{National Research Nuclear University ``MEPhI'' (Moscow Engineering Physics Institute), 115409 Moscow, Russia}
\author{H.~Ruhl}\affiliation{Ludwig-Maximilians Universit\"{a}t, M\"unchen, 80539, Germany}
%\date{}

\begin{abstract}
The concept of ponderomotive potential is upgraded to a regime in which radiation friction becomes dominant. The radiation friction manifests itself in novel features of long-term capturing of the particles released at the focus and impenetrability of the focus from the exterior. We apply time scales separation to the Landau-Lifshitz equation splitting the particle motion into quivering and slow drift of a guiding center. The drift equation is deduced by averaging over fast motion.
\end{abstract}

\maketitle

A concept of ponderomotive potential \cite{Gaponov:1958} had been proven to be an extremely useful heuristic tool for interpreting electron scattering in experiments with focused laser fields, see e.g. \cite{Bucksbaum:1987,Hartemann:1995,Salamin:1997,Meyerhofer:1999}. Generalizations to relativistic intensities are straightforward but 
technically more tedious \cite{Kibble:1966,Bauer:1995,Narozhny:2000}. However, none of the papers \cite{Gaponov:1958,Bucksbaum:1987,Hartemann:1995,Salamin:1997,Meyerhofer:1999,Kibble:1966,Bauer:1995,Narozhny:2000} took into account effect of radiation friction because of relatively low laser intensity available at that time. The intensity of recent laser sources has been increased up to $2\times 10^{22}$W/cm$^2$ \cite{Yanovsky:2008}, thus enabling entering into ultra-relativistic regime for which the effects of radiation friction could become highly significant (see \cite{Rorlich:2007,Macchi:2012,DiPiazza:2012} for an overview). The unusual behavior of electrons in ultra-strong fields was recently revealed in numerical simulations \cite{Gonoskov:2013,Pukhov:2014,Elkina:2014}. However, none of these examples have yet convincing physical explanation. The purpose of this Letter is to analyze general properties of electron motion under the dominance of radiation friction as well as to outline a unifying framework of time scale separation leading to guiding center equations of motion for radiative electrons. As a result, we develop a new analytical approach which allows to formulate, explore and explain long-time radiative capturing of particles inside the focus and impenetrability of the focus for particles from the exterior.

A classical particle motion with radiation friction is governed by the Landau-Lifshitz equation \cite{LL,Rorlich:2007}, where we retain just the leading term of radiation friction force $\vec{F}_R$ proportional to the energy squared \footnote{We use units $\hbar=c=1$ throughout the Letter.}:
\begin{align}
&\frac{d\vec{p}}{dt}=\vec{F}_L+\vec{F}_R,\quad
\vec{F}_L=e\left(\vec{E}+\vec{v}\times\vec{B}\right),\label{LL_eq}\\
&\vec{F}_R\approx-\frac23\frac{e^4}{m^4}\left[\left(\varepsilon\vec{E}+\vec{p}\times\vec{B}\right)^2-\left(\vec{p}\cdot\vec{E}\right)^2\right]\vec{v},\nonumber
\end{align}
where $\vec{p}$, $\varepsilon(\vec{p})=\sqrt{p^2+m^2}$ and $\vec{v}=\vec{p}/\varepsilon$ are the momentum, energy and velocity of a particle, respectively. Note that radiation friction dominates over the Lorentz force if the dimensionless field strength $a_0=\frac{e}{m}\sqrt{\left\langle {\vec{A}}^2\right\rangle}$ exceeds the critical value $a_{0*}=\left(3m/2e^2\omega\right)^{1/3}\simeq 470$, where the numerical value is given for optical frequency $\hbar\omega\simeq 1\text{eV}$ \cite{Bulanov:2004,DiPiazza:2012}. This roughly corresponds to laser intensity $I\gtrsim 5\times 10^{23}$W/cm$^2$. In what follows we assume that the field is that strong, $a_0\gtrsim a_{0*}$.

We also assume that the field is weakly inhomogeneous of the form 
$\vec{A}(\mu\vec{r},t)$, where $\mu\ll 1$ is a small parameter.  A 2D field configuration of this kind can be realized in a focal plane ($z=0$) of a couple of counter-propagating identically shaped focused monochromatic laser beams. In this context $\mu$ is the angular aperture of the pulses, $w_0=(\omega\mu)^{-1}$ is their waist and $z_R=(2\omega\mu^2)^{-1}$ is the Rayleigh length. Note that at the focal plane the electric field $\vec{E}=-\partial\vec{A}/\partial t=O(1)$, while the magnetic field $\vec{B}=\nabla\times\vec{A}$ is $O(\mu)$.

We split the particle motion $\vec{r}(t)=\mu^{-1}\vec{R}(\mu^2 t)+\vec{\xi}(t)$ into a slow drift of a guiding center and rapidly oscillating motion with small amplitude around it. Both functions can be expanded in powers of $\mu$, $\vec{R}\approx\vec{R}_0+\mu\vec{R}_1$, $\vec{\xi}\approx\vec{\xi}_0+\mu\vec{\xi}_1$. Accordingly, $\vec{r}\approx\mu^{-1}\vec{r}_0+\vec{r}_1$, $\vec{v}\approx\vec{v}_0+\mu\vec{v}_1$ and $\vec{p}\approx\vec{p}_0+\mu\vec{p}_1$, where $\vec{r}_0=\vec{R}_0(\mu^2 t)$, $\vec{v}_0=\vec{\xi}_0'(t)$ and 
\begin{align}
\vec{r}_1=\vec{\xi}_0(t)+\vec{R}_1(\mu^2 t),\quad  \vec{v}_1=\vec{\xi}_1'(t)+\vec{R}_0'(\mu^2 t).\label{expand}
\end{align}

Then, substituting these expansions into Eq.~(\ref{LL_eq}) and equating the terms of the same powers of $\mu$, we obtain
\begin{widetext}
\begin{align}
\dot{\vec{p}}_0=&e\vec{E}_0(t)-\frac23\frac{e^4}{m^4}\left[\varepsilon_0^2 E_0^2-\left(\vec{p}_0\cdot\vec{E}_0(t)\right)^2\right]\vec{v}_0,\label{p0}\\
\dot{\vec{p}}_1=&e\left(\vec{E}_1+\vec{v}_0\times\vec{B}_1\right)-\frac23\frac{e^4}{m^4}\left\{\left[\varepsilon_0^2E_0^2-\left(\vec{p}_0\cdot\vec{E}_0\right)^2\right]\vec{v}_1+2\varepsilon_0E_0^2\left(\vec{v}_0\cdot\vec{p}_1\right)\vec{v}_0-2\left(\vec{p}_0\cdot\vec{E}_0\right)\left(\vec{E}_0\cdot\vec{p}_1\right)\vec{v}_0+
\right.\nonumber\\ 
&\left.+2\varepsilon_0^2\left(\vec{E}_0\cdot\vec{E}_1\right)\vec{v}_0-2\varepsilon_0\left[\left(\vec{E}_0\times\vec{B}_1\right)\cdot\vec{p}_0\right]\vec{v}_0-2\left(\vec{p}_0\cdot\vec{E}_0\right)\left(\vec{p}_0\cdot\vec{E}_1\right)\vec{v}_0\vphantom{\frac12}\right\},\label{p1}\\
\vec{v}_1=&\frac1{\varepsilon_0}\left[\vec{p}_1-\left(\vec{v}_0\vec{p}_1\right)\vec{v}_0\right],\label{r1}
\end{align}
\end{widetext}
where $\vec{v}_0=\vec{p}_0/\varepsilon_0$, $\varepsilon_0=\varepsilon(\vec{p}_0)$, $\vec{E}_0=\vec{E}(\vec{r}_0,t)$, $\vec{E}_1=(\vec{r}_1\cdot\nabla)\vec{E}(\vec{r}_0,t)$ and $\vec{B}_1=\vec{B}(\vec{r}_0,t)$. 

Now suppose that the vector $\vec{E}_0(t)=\vec{E}(\vec{r}_0,t)$ is lying in the plane $z=0$ and at each fixed $\vec{r}_0$ is rotating uniformly around the $z$-axis with angular velocity $\vec{\omega}$, while the magnetic field is oscillating along $z$-axis (this implies circular polarization of the colliding beams). In such a case the zeroth order equation (\ref{p0}) admits an exact solution \cite{Zeldovich:1975}. Namely, after transition into the reference frame rotating along with the field this equation becomes autonomous, with an attractor, for which the momentum $\vec{p}_0$ is lagging behind the field by an angle $\delta$. This angle, together with the magnitude of the momentum, are determined by the equations
\begin{align}
\cos{\delta}=\left(a_0/a_{0*}\right)^3\sin^4{\delta},\quad p_0=ma_0\sin{\delta}.\label{Zeld}
\end{align}
If $a_0\ll a_{0*}$, then radiation friction is negligible and $\delta\approx\pi/2$. In the opposite regime of dominance of radiation friction ($a_0\gtrsim a_{0*}$) $\delta$ is small and (\ref{Zeld}) can be reduced to $\delta\approx (a_{0*}/a_0)^{3/4}\ll 1$, $p_0\approx m(a_{0*}^3a_0)^{1/4}$. The settling time $\tau_s\sim \delta/\omega$ is estimated  from linearized near the attractor Eq.~(\ref{p0}). For $\delta\ll 1$ it is much smaller than the rotation period and this is one of the typical features of the regime with dominating radiation friction. This essentially means that strong radiation friction reduces the effective number of degrees of freedom at time scales larger than rotation period.

Next we turn to consideration of the first order equations (\ref{p1}), (\ref{r1}). As follows from (\ref{expand}), the corrections $\vec{r}_1$, $\vec{v}_1$ and $\vec{p}_1$ contain both slow and fast motions, $\vec{p}_1=\vec{p}_1^{(s)}(\mu^2t)+\vec{p}_1^{(f)}(t)$. In our case, the first order correction $\vec{\xi}_1(t)$ to fast motion represents rotation with doubled frequency $2\omega$. To separate slow and fast motions, we consider balance of the second Fourier harmonics in Eq.~(\ref{p1}) to find $\vec{p}_1^{(f)}$, then substitute it back and after averaging over a rotation period obtain $\vec{p}_1^{(s)}$. Finally, by substituting the results into Eq.~(\ref{r1}) and averaging, we obtain the equation
\begin{align}
\frac1{\mu}\frac{d\vec{R}_0}{dt}\approx -\frac{\mu}{2\omega a_0(\vec{R}_0)}\left[\vec{n}_\omega\times \nabla a_0(\vec{R}_0)+\frac54\delta \nabla a_0(\vec{R}_0)\right],\label{drift}
\end{align}
where $\vec{n}_\omega$ is a unit vector directed along $\vec{\omega}$ (i.e., the focal axis). For the sake of clarity, we present here Eq.~(\ref{drift}) in a simplified form, with the accuracy $O(\delta^2)$ \footnote{Note that $60\%$ contribution to the value of the numerical coefficient of the second term comes from coupling with second harmonic. It can be shown that higher even order harmonics are also present, but are much smaller and in our case slow motion is decoupled from them.}. 

\begin{figure}[th!]
\includegraphics[width=\columnwidth]{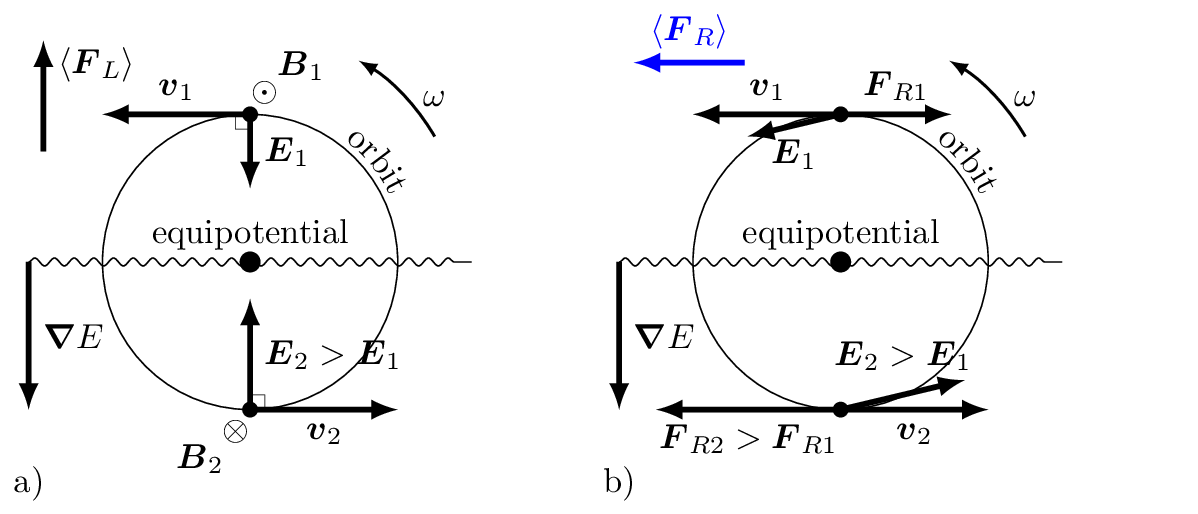}
\caption{\label{force_balance}
(Color online) Balance of forces acting on a single loop path of fast motion in inhomogeneous field: a) case of relativistic ponderomotive effect ($a_0\ll a_{0*}$): both the Lorentz force $\vec{v}_0\times\vec{B}_1$ and the resultant of electric forces acting on top and bottom of a loop are directed upwards, i.e. opposite to the gradient of the field amplitude $\nabla a_0$; b) case of dominance of radiation friction ($a_0\gg a_{0*}$): the resultant of radiation friction forces on top and bottom of the loop is directed almost to the left, i.e. opposite to $\vec{n}_\omega\times\nabla a_0$.}
\end{figure}

In a sharp contrast to the ponderomotive effect regime for moderate $a_0\lesssim a_{0*}$, the slow drift motion is governed by a first order equation instead of the ponderomotive potential (i.e. second order equation). This is a principle feature of the regime of dominance of radiation friction and is attributed to the aforementioned reduction of the degrees of freedom due to presence of a strong friction. The same degeneration takes place e.g. in an inhomogeneous magnetic field, although due to a different physical reason of energy conservation in magnetic field \cite{vanKampen:1985}.

Interestingly, in virtue of Eq.~(\ref{drift}), drift is directed almost along the equipotentials $a_0=\text{const}$, while ponderomotive force usually pulls the particles across the equipotentials of the ponderomotive potential $U_\text{pond}=m\sqrt{1+a_0^2}$. Thus, the actions of ponderomotive force in the regimes $a_0\gtrsim a_{0*}$ and $a_0\lesssim a_{0*}$ are almost orthogonal to each other. This feature is illustrated at a qualitative level in Fig.~\ref{force_balance}b), which explains why the resultant force acting on a single loop part of the trajectory $\vec{r}=\vec{\xi}_0(t)$ in an inhomogeneous field is directed orthogonally to the gradient of the field amplitude. This can be compared to the force balance in Fig.~\ref{force_balance}a), which illustrates just the usual ponderomotive effect. However, for finite values of $a_0$, due to a smaller term $\propto\delta$ originating from the Lorentz force, the drift still possesses a small component directed outwards. Only in the formal limit $a_0\to\infty$ this component vanishes and particle drifts strictly along the equipotential. 

\begin{figure}[th!]
\begin{center}
\includegraphics[width=0.9\columnwidth]{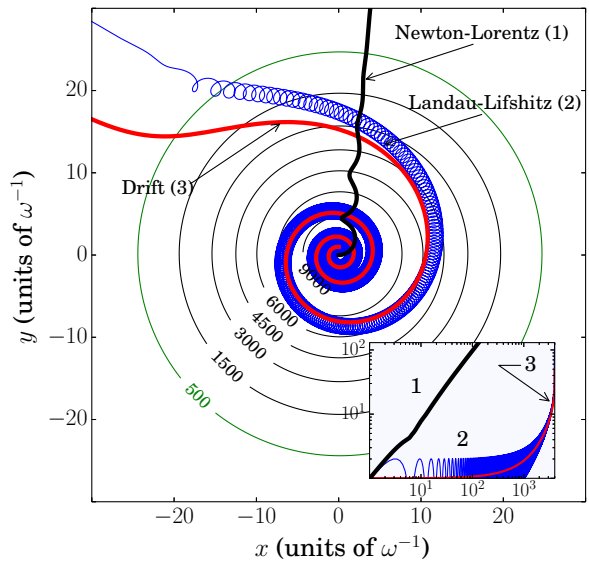}
\end{center}
\caption{\label{trajectories}
(Color online) The numerical solutions of the Newton-Lorentz equation (with radiation friction neglected, curve 1), the Landau-Lifshitz equation (curve 2) and the solution of the drift equation Eq.~(\ref{drift}) (curve 3) combined together with the levels of $a_0$. The employed field model is $E_x=(m\omega/e) a_0(r)\cos{\omega t}$, $E_y=(m\omega/e)a_0(r)\sin{\omega t}$, $B_z=(m/e) a_0'(r)\cos(\omega t -\phi)$ (other field components vanish in the $xy$ plane), where $a_0(r)=a_{0m}\exp(-\mu^2\omega^2r^2/2)$, $a_{0m}=10^4$ and $\mu=0.1$. Particle starts from the origin. Inset: dependence $r(t)$ (all in units of $\omega^{-1}$) for the same solutions.}
\end{figure}

This is demonstrated in Fig.~\ref{trajectories} for a Gaussian profile of the beam. In order to emphasize the effect, we have chosen very high $a_0=10^4$ and $\mu=0.1$. The numerical solution of the Landau-Lifshitz equation (curve 2), is compared to the solution without radiation friction (curve 1) and the untwisting spiral 3, which corresponding to the solution of the drift equation Eq.~(\ref{drift}). One can immediately observe a drastically different behavior of curves 1 and 2. At the same time, curve 3 approximates correctly the drift of a guiding center in the time range $t\lesssim \mu^{-2}\omega^{-1}$, for which $r\sim\mu^{-1}R_0\lesssim w_0$. Further divergence of the curves 2 and 3 is explained by the corrections $O(\mu^2)$ not considered here. Right after intersecting the level $a_0=a_{0*}$ the curve 2 stops circling and continues like the curve 1, as expected. For smaller values of $a_0\gtrsim a_{0*}$ the number of windings of the trajectory around the focus is of course smaller, but the whole picture looks much the same.

Because of strong Lorentz time contraction in the lab frame, the time scales of quivering and drift for the curve 1 for our choice of parameters look commensurable. But with radiation friction taken into account, the drift remains non-relativistic in the time range $t\lesssim \mu^{-2}\omega^{-1}$, no matter how strong is the field, just because the drift velocity in Eq.~(\ref{drift}) is $O(\mu)$. This point is illustrated at the inset in Fig.~\ref{trajectories}, where distance from the origin $r(t)$ is depicted with and without account for radiation reaction. As a result, the trajectory 2 indeed demonstrates a nice separation of time scales thus proving applicability of time scales separation technique. But it also means that straightforward solution of the Landau-Lifshitz equation becomes a challenge for numerical methods in use. In order to gain the desired resolution and accuracy at an acceptable expense, we apply the approach suggested in Ref.~\cite{Elkina:2014}, which is based on application of implicit schemes to the covariant equations of motion formulated in a proper time.
 
\begin{figure}[th!]
\begin{center}
\includegraphics[width=0.9\columnwidth]{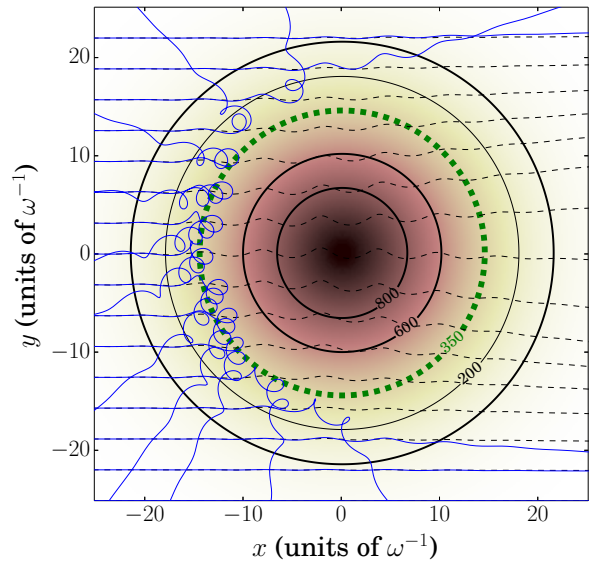}
\end{center}
\caption{\label{scattering}
(Color online) Trajectories of particles scattered from the same Gaussian-shaped field as in Fig.~\ref{trajectories} for different values of impact parameter with (solid lines) and without (dashed lines) account for radiation friction ($\gamma_i=2\times 10^3$, $a_{0m}=10^3$, $\mu=0.1$). Numbers mark the levels of $a_0(r)$.}
\end{figure}

Another, and even more general, feature of the radiation friction dominated regime is impenetrability of the core of a localized field for high energy particles incoming from the exterior. This effect admits a rather simple explanation in terms of Pomeranchuk bound for radiative cooling \cite{Pomeranchuk:1939,LL}. This bound was introduced in order to explain energy losses by cosmic rays in the earth magnetosphere. The energy of a particle coming from the exterior with initial gamma-factor $\gamma_i\gg (a_{0*}^3/a_0)^{1/2}$ after a long enough path (in our case $\Delta t\sim w_0/2\gg\omega^{-1}$) in the field is reduced due to radiative cooling down to the value bounded by
\begin{align}
\gamma_f<\gamma_f^{(max)}=\frac{3m^3/2e^2}{\int\limits_{t_i}^{t_f}F_{L\perp}^2(\vec{r}(t),t)\,dt}\sim
\frac{4 a_{0*}^3\mu}{a_0^2}\ll \gamma_i,\label{Pom_b}
\end{align}
where $\vec{F}_{L\perp}$ is a component of the Lorentz force transverse to the velocity.
After cooling is completed,  as a qualitative estimate, we can apply a criterion $\gamma_f>a_0$ for penetrability of a ponderomotive potential hill (thus neglecting radiation friction at this stage). But it is evident that such a criterion, being combined with Eq.~(\ref{Pom_b}), means impenetrability of the core of the field with respect to arbitrarily high energy particles if $a_0\gtrsim a_{0*}(4\mu)^{1/3}$. For example, for $\mu=0.1$ the threshold value of $a_0$ would be of the order of $350$. These estimates are in perfect agreement with numerical calculations. Indeed, all the scattering particles are reflected outwards just around the pointed out equipotential level, see Fig.~\ref{scattering}. The 2D consideration in the focal cross section can be readily extended 
into 3D, where field satisfying Maxwell equations must possess a node-antinode structure along the focal axis. Then the high-energy particle crossing the field may partially penetrate through the nodes. However, the core regions surrounding antinodes still remain impenetrable for high-energy particles even in a 3D setup.

The results obtained within classical treatment based on the Landau-Lifshitz equation (\ref{LL_eq}) have to be taken with certain care since at the specified extreme values of parameters quantum nature of radiation could be also important. However, quantum simulations based on the Monte-Carlo algorithm described in Ref.~\cite{Elkina:2011} reveal qualitatively the same behavior. The only important difference is that, due to random quantum recoils from photon emissions, the sampled trajectories become stochastic, with a noticeably wide spread around the classical trajectories Ref.~\cite{DiPiazza:2013}. In order to explain the origin of qualitative accordance of classical and quantum simulations, we estimate the dynamical quantum parameter (proper acceleration in Compton units) $\chi\simeq F_{L\perp}\gamma/m^2$ for parameters under consideration. For a particle released at the origin and $a_0=10^4$ we have $\delta\approx\mu=0.1$, so that the components $E\delta$, $B\sim\mu E$ of the electric and magnetic fields transverse to velocity are of the same order. Then, taking into account Eq.~(\ref{Zeld}), we can estimate $\chi\sim (2\omega/m)a_0^2\delta^2\sim 4$. Since it exceeds unity but not too much, quantum effects such as the spread of sampled trajectories must be indeed noticeable though not changing the whole picture completely. On the other hand, in a scattering setup the parameter $\chi$ remains even smaller than unity, attaining its maximum during radiative cooling, as it goes faster than particle approaches the core of the field. After cooling, the particles in fact behave classically. Nevertheless, quantum spread shows up in this setup as well because quantum fluctuations are amplified by the billiard instability of scattering dynamics.

To conclude, by analysis of the motion of charged particles in a 2D weakly inhomogeneous circularly polarized fields in radiation friction dominated regime, we reveal three rather basic features which discriminate this regime against the regime of usual motion in ponderomotive potential: (i) The trajectories of particles set in an equilibrium regime after a short time less than a period. In this regime the motion is well separated into a fast quiver motion and a drift of its guiding center. This drift is directed almost along the levels of the field amplitude, constituting with them a small angle $\delta$. This is a principle difference with ponderomotive pulling across the levels of ponderomotive potential; (ii) Even more importantly, the drift velocity (\ref{drift}) is proportional to $\mu$, making  drift non-relativistic [the component of the drift velocity across the levels of $a_0$ is even more suppressed by a factor $\delta$ because of (i)]. Combined together, (i) and (ii) mean that a particle placed initially inside the core of the field, would be long-term captured inside. This is in great contrast to the case of weaker fields, for which it would be rapidly pushed away with almost the speed of light from the top of a ponderomotive potential hill. (iii) Finally, we predict existence of a threshold field strength, above which any focused field must become completely impenetrable for high-energy particles. Our predictions (i)-(iii) could be tested with the forthcoming ELI and/or XCELS projects \cite{ELI,XCELS}. More sophisticated experiments could be proposed for probing the internal structure of focused fields, since more elaborate spatial envelope profiles of the laser pulses with localized wells may allow to capture radiatively cooled fractions of particles and trap them inside over laser pulse duration. This sort of plasma trapping can have potentially a number of useful applications, including design of new $\gamma$-ray sources \cite{Bashinov:2013} and stimulating production of self-sustained QED cascades \cite{Elkina:2011}.

\begin{acknowledgments}
This research was supported by the Russian Fund for Basic Research (grant 13-02-00372), the RF President program for support of young Russian scientists and leading research schools (grant NSh-4829.2014.2), the Sonderforschungsbereiche TransRegio-18 (Project B13), the Cluster-of-Excellence Munich-Centre for Advanced Photonics (MAP), and Arnold Sommerfeld Zentrum (ASZ). 
\end{acknowledgments}

\end{document}